\documentclass[12pt]{emulateapj}
\usepackage{graphicx}

\newcommand{\RA}[3]{\mbox{R.A.}={#1}$^{{\rm h}}${#2}$^{{\rm m}}${#3}$^{{\rm s}}$}
\newcommand{\decl}[3]{\mbox{Dec.}={#1}$^{\circ}${#2}\arcmin{#3}\arcsec}

\newcommand{\swift}{\emph{Swift}}
\newcommand{\chisq}{$\chi^2$}
\newcommand{\cm}[1]{~cm$^{#1}$}

\newcommand{\cts}{~cts\,s$^{-1}$}
\newcommand{\e}[1]{10$^{#1}$}
\newcommand{\ee}[1]{$\times$10$^{#1}$}

\shorttitle{\swift~observations of GRB070110}
\shortauthors{E. Troja et al.}

\begin{document}
\title{{\it Swift} observations of GRB~070110: 
an extraordinary X-ray afterglow powered by the central engine.}

\author{E.~Troja\altaffilmark{1,2,3}, G.~Cusumano\altaffilmark{1}, P.~O'Brien\altaffilmark{2},
B.~Zhang\altaffilmark{4}, B.~Sbarufatti\altaffilmark{1}, V.~Mangano\altaffilmark{1}, 
R.~Willingale\altaffilmark{2}, G.~Chincarini\altaffilmark{5,6}, J.~P.~Osborne\altaffilmark{2}, 
F.~E.~Marshall\altaffilmark{7}, D.~N.~Burrows\altaffilmark{8}, S.~Campana\altaffilmark{6}, N.~Gehrels\altaffilmark{7}, 
C.~Guidorzi\altaffilmark{5,6}, H.~A.~Krimm\altaffilmark{7,9}, V.~La~Parola\altaffilmark{1}, E.~W.~Liang\altaffilmark{4,10},
T.~Mineo\altaffilmark{1}, A.~Moretti\altaffilmark{6}, K.~L.~Page\altaffilmark{2}, P.~Romano\altaffilmark{5,6}, 
G.~Tagliaferri\altaffilmark{6}, B.~B.~Zhang\altaffilmark{4,11}, M.~J.~Page\altaffilmark{12}, P.~Schady\altaffilmark{12}
}

\altaffiltext{1}{INAF - Istituto di Astrofisica Spaziale e Fisica Cosmica,
                 Sezione di Palermo, via Ugo la Malfa 153, 90146 Palermo, Italy}
\altaffiltext{2}{Department of Physics and Astronomy, University of Leicester,
                 Leicester, LE1~7RH, UK}
\altaffiltext{3}{Dipartimento di Scienze Fisiche ed Astronomiche,
                 Sezione di Astronomia, Universit\`a di Palermo, 
		 Piazza del Parlamento 1, 90134 Palermo, Italy}
\altaffiltext{4}{Department of Physics and Astronomy, University of Nevada Las Vegas, NV, USA}
\altaffiltext{5}{Universit\`a degli studi di Milano-Bicocca,
                 Dipartimento di Fisica, piazza delle Scienze 3, 
                 I-20126 Milano, Italy}
\altaffiltext{6}{INAF -- Osservatorio Astronomico di Brera, 
                 via Emilio Bianchi 46, I-23807 Merate (LC), Italy}
\altaffiltext{7}{NASA, Goddard Space Flight Center, Greenbelt, MD 20771, USA}		 
\altaffiltext{8}{Department of Astronomy \& Astrophysics, 525 Davey Lab., 
                 Pennsylvania State University, University Park, PA 16802, USA}
\altaffiltext{9}{Universities Space Research Association, 10211 Wincopin Circle, 
                 Suite 500, Columbia, MD 20144 USA}
\altaffiltext{10}{Department of Physics, Guangxi University, Nanning 530004, China}
\altaffiltext{11}{National Astronomical Observatories/Yunnan Observatory, CAS, 
                  Kunming 650011, China}
\altaffiltext{12}{UCL Mullard Space Science Laboratory, Holmbury St Mary, Dorking, Surrey RH5 6NT}

\begin{abstract}
We present a detailed analysis of \swift~multi-wavelength observations of GRB~070110 and its 
remarkable afterglow. The early X-ray light curve, interpreted as the tail of the prompt emission,
displays a spectral evolution already seen in other gamma-ray bursts.
The optical afterglow shows a shallow decay up to $\sim$2~d after the burst, which is
not consistent with standard afterglow models.
The most intriguing feature is a very steep decay in the X-ray flux at $\sim$2\ee{4}~s after the burst, 
ending an apparent plateau. The abrupt drop of the X-ray light curve rules out 
an external shock as the origin of the plateau in this burst and implies long-lasting activity of the central engine. The temporal and spectral properties
of the plateau phase point towards a continuous central engine emission rather than the episodic emission of X-ray flares.
We suggest that the observed X-ray plateau is powered by a spinning down central engine, 
possibly a millisecond pulsar, which dissipates energy at an internal radius before 
depositing energy into the external shock.

\end{abstract}

\keywords{gamma rays: bursts; X-rays: individual (GRB~070110)}

\section{Introduction}\label{sec:intro}

The Swift Gamma Ray Burst Explorer \citep{swift04} 
is a multi-wavelength observatory specifically designed
to study gamma-ray burst (GRB) evolution from their early stages.
It is equipped with a wide-field instrument, the Burst Alert Telescope
(BAT; \citealt{bat05}), covering the 15-350 keV energy band,
and two narrow field instruments, the X-Ray Telescope (XRT; \citealt{xrt05}) 
and  the Ultraviolet/Optical Telescope (UVOT; \citealt{uvot05}),
covering the 0.2-10 keV band and the 1700-6500~\AA\ wavelength range,
respectively.

During the first two years of its mission \swift\ detected $\sim$200 bursts,
providing well-sampled X-ray and optical afterglow light curves.
\citet{nousek06} and \citet{zhang06} identified a common pattern in the X-ray light curves,
described as consisting of three power law segments: 
an early steep decay with a temporal slope 3$\leq$$\alpha$$\leq$5 lasting up to $\sim$300~s, 
followed by a shallower phase (0.5$\leq$$\alpha$$\leq$1);
the slope of the light curve steepens again (1.0$\leq$$\alpha$$\leq$1.5) 
at $\sim$\e{3}-\e{4}~s after the burst onset.
X-ray flares, overlaid on the underlying power law decay, have been detected up to $10^5$~s
after the BAT trigger \citep{burrows05,chincarini07}.

\citet{obrien06} modeled the observed shape of early X-ray light curves
with an exponential decay that relaxes into a power law and, in most cases,
presents the shallow phase. They showed that
the initial rapid decay is a smooth extension of the prompt emission, 
probably due to emission from large angles relative to the observer's line of
sight \citep{kumar00}.
A long-lasting energy injection into the forward shock, 
due either to a late internal activity or to a radial distribution of Lorentz factors,
has been invoked to interpret the shallower stage of the X-ray light curve
\citep{remes98,zhang02,granot06}.
This explanation is consistent with the final smooth steepening 
into the ``standard afterglow'' decay \citep{mesre97}.

The afterglow of GRB~070110 cannot be easily traced back to this well-known scenario.
Its early X-ray light curve seems to display a canonical shape, 
with an initial steep decay and then a very flat plateau phase ($\alpha\sim$0.05),
but $\sim$2\ee{4}\,s after the trigger the count rate drops abruptly by more 
than one order of magnitude with a slope $\alpha$$>$7 \citep{gcn6008,report}. 
The behavior of the optical afterglow seen by the UVOT 
is very different, showing a shallow smooth decay 
remaining fairly bright at later times.

In the light of the unique properties of its afterglow, the \swift~team
identified GRB~070110 as a  `burst of interest' encouraging a 
follow-up campaign \citep{gcn6014}.

The optical afterglow of GRB~070110 \citep{report}
was also detected 
by the ESO VLT equipped with the FORS2 instrument.
A redshift of $z$=2.352$\pm$0.001 has been inferred
on the basis of several absorption features in the spectra \citep{gcn6010}.
Further VLT observations, performed $\sim$10.7~d after the burst, 
detected a fainter than predicted afterglow, 
suggesting the presence of an optical break \citep{gcn6021}.

In this paper we report on the $\gamma$-ray, X-ray and
optical observations performed by \swift.
The paper is organized as follows: 
in \S~2 we present a multiwavelength timing and spectral analysis
of both the prompt and the afterglow emission; 
in \S~3 we discuss our results.
Finally, in \S~4 we summarize our findings and conclusions.

Throughout the paper, times are given relative
to the BAT trigger time T$_0$, t=T-T$_0$,
and the convention F$_{\nu,{\rm t}} \propto \nu^{-\beta}{\rm t}^{-\alpha}$
has been followed, where the energy index $\beta$ is related to the
photon index $\Gamma=\beta+1$.
We have adopted a standard cosmology model with Hubble constant
H$_0$=70\,km\,s$^{-1}$\,Mpc$^{-1}$ and cosmological constants
$\Omega_\Lambda=0.73$, $\Omega_M=0.27$ \citep{wmap}. 
The phenomenology of the burst is presented in the observer time
frame, unless otherwise stated.

All the quoted errors are given at 90\% confidence level for one interesting
parameter ($\Delta$\chisq=2.706, \citealt{lampton76}).

\section{Data analysis}\label{sec:data}

\subsection{Observations}
GRB~070110 triggered the \swift~BAT at 07:22:41 UT on 10th January, 2007.
The \swift~narrow field instruments, XRT and UVOT, began observing 93~s and 85~s after the trigger,
respectively. An accurate afterglow position was rapidly determined by the UVOT at
\RA{00}{03}{39.20}, \decl{-52}{58}{26.3} (J2000, \citealt{report}), with an uncertainty of 1~arcsec.

XRT observations began with an initial 2.5~s image mode frame and then, 
as the source was bright ($\sim$40\cts), collecting data in  Window Timing (WT) mode.
The XRT automatically switched to Photon Counting (PC) mode when the source
decreased to $\sim$2\cts.
Follow-up observations lasted 26 days for a total net exposure of 165 s in WT mode 
and 330 ks in PC mode.

The UVOT took a short exposure with the $V$ filter while the spacecraft was settling at the end of the initial slew.
This exposure was followed by a ``finding chart'' exposure with the White filter lasting 100 s
and then an exposure with the $V$ filter lasting 400 s. 
UVOT then began its usual procedure of cycling through its 3
visible filters ($V$, $B$, and $U$) and 3 $UV$ filters ($UVW1$, $UVM2$, and $UVW2$). 
The optical afterglow was detected in the White, $U$, $B$, and $V$ filters, but not in the $UV$ 
filters. 
The lack of detection of the afterglow in the $UV$ filters is consistent with the measured redshift.
A total of 219 UVOT exposures were taken in the first 6.7 days,
after which the afterglow fell below the detection thresholds.

\subsection{Gamma-ray data}\label{sec:bat}

\begin{figure}
\centering
\includegraphics[angle=270,scale=0.35]{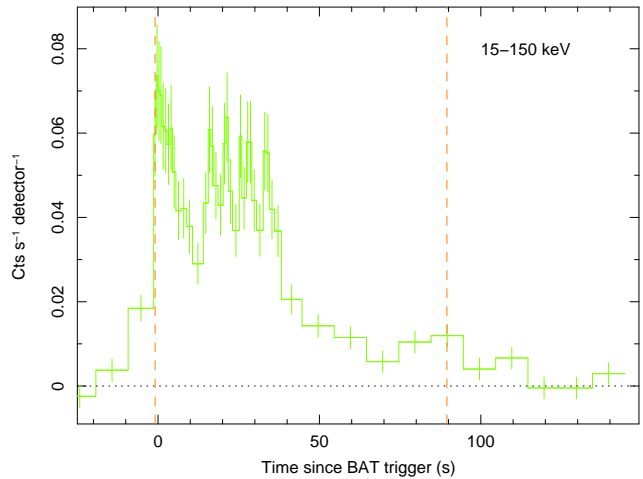}
\caption{BAT Mask-weighted light curve (15--150~keV) of GRB~070110. 
Dashed vertical lines mark the T$_{90}$ duration.}
\label{fig:batlc}
\end{figure}

We analyzed BAT event data using the standard BAT analysis software 
included in the NASA's HEASARC software (HEASOFT, version~6.1.2).
Fig.~\ref{fig:batlc} presents the BAT mask-weighted light curve in the 15-150 keV energy band.
It shows a first main peak at t$\sim$0~s and then a decay on 
which several peaks are superposed; emission is visible until t$\sim$100~s.
We estimated the burst duration, defined as the interval containing  
90\% of the total observed fluence, to be T$_{90}$ (15-150~keV)=89$\pm$7~s.

Spectra were created using the task {\tt batbinevt}, updating relevant keywords with 
{\tt batupdatephakw}.
The corresponding response matrices were generated by the task {\tt batdrmgen}. 
Systematic errors were properly added to the spectra.
Since the spacecraft slew started $\sim$40~s after the trigger, 
we created two different spectra (before and during the slew) with the appropriate response
matrices and performed a joint fit.

The time-averaged T$_{90}$ spectrum (from -0.4~s to 89~s) can be fitted with a simple power
law of photon index 1.57$\pm$0.12. A cut-off power law or a Band model \citep{band93} do not provide a better description
and cannot constrain the peak energy value. 
In the hardness ratio light curves, comparing different BAT energy bands,
no sign of a significant spectral evolution is present throughout the prompt emission.

The fluence over the 15-150 keV band is (1.8$^{+0.2}_{-0.3}$)\ee{-6}~erg~cm$^{-2}$,
from which we derive an observed isotropic energy of 2.3\ee{52}~erg.
This value can be considered as a lower limit to the isotropic energy, 
E$_{\gamma,\rm iso}$, which is defined over a larger energy band (1~keV-10~MeV in the source rest frame).
Extrapolating our best fit model over the whole 1~keV-10~MeV (source rest frame)
we estimate an upper limit of E$_{\gamma,\rm iso}$$\leq$1.3\ee{53}~erg.
This value is derived under the extreme assumption that the best description
of the spectrum is Band law \citep{band93} with a peak energy E$_{p}>$10~MeV in the source rest frame.


\begin{figure*}
\centering
\includegraphics[angle=270,scale=0.55]{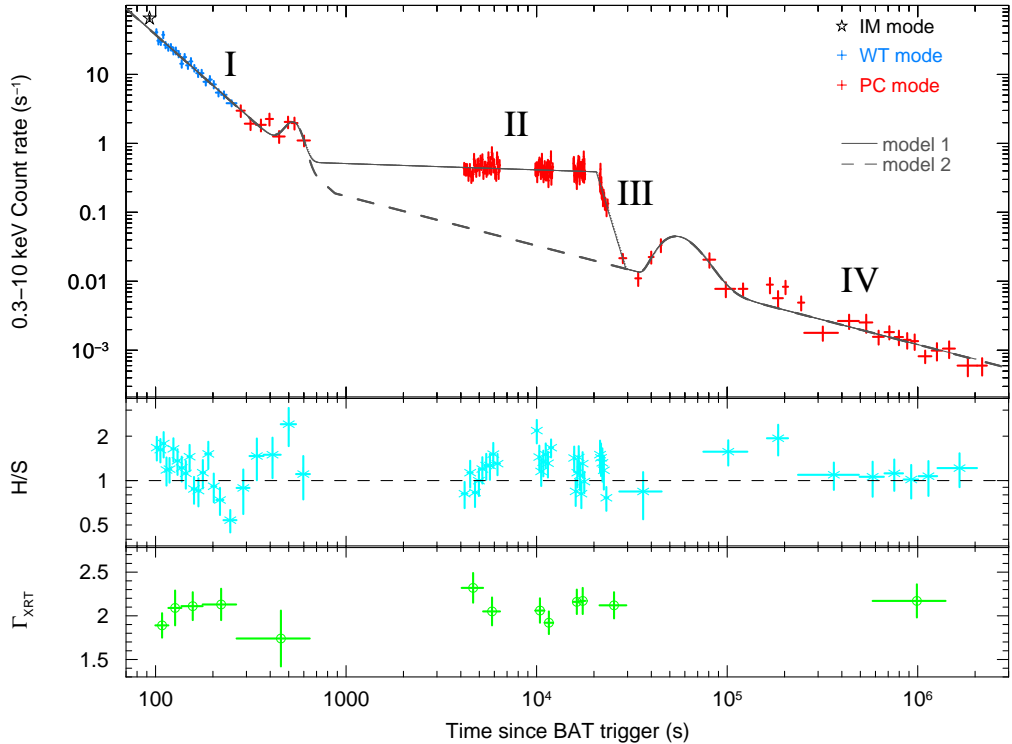}
\caption{{\it Upper panel:} XRT light curve of GRB070110 in the 0.3-10~keV band.
The two models described in the text are shown: 
power law segments with three temporal breaks ({\it solid line})
and a simple broken power law ({\it dashed line}). The bump at t$\sim$530~s
is modeled with a Gaussian function, the late one at t$\sim$5\ee{4}s with a FRED profile.
The four phases of the X-ray light curve are marked: (I) an early decay, (II) an apparent plateau
followed by (III) a rapid drop, and (IV) a final shallow decay.
{\it Middle panel:} Hardness ratio (H/S) light curve. It compares 
source counts in the hard band (H: 1-10~keV) and in the soft band (S: 0.3-1~keV).
{\it Lower panel:} Photon index $\Gamma_{\rm XRT}$ temporal variations. These values were derived
fitting the X-ray spectra with an absorbed power law model. 
}
\label{fig:xrtlc}
\end{figure*}


\subsection{X-ray data}\label{sec:xray}

XRT data were processed using the XRTDAS software package (v.~2.0.1)
distributed within HEASOFT. We used the latest release of the XRT
Calibration Database (CALDB~2.6) and applied standard screening criteria 
to obtain the final cleaned event list.
We selected only events with grades 0-12 for PC mode data and 0-2 for WT mode data.
Such a selection provides the best combination of spectral resolution and detection efficiency.
Our analysis has been performed over the 0.3-10 keV energy band.

\subsubsection{Temporal analysis}\label{sec:lc}

We extracted WT data in a rectangular region, 40$\times$20 pixels wide,
centered on the source position. The background contribution was estimated from a region,
with the same shape and size, sufficiently offset ($>$2 arcmin) from the source position to avoid
contamination from the PSF wings and free from contamination by other sources.

The first 300~s of PC observations were affected by pile-up. To account for this,
we chose an annular extraction region centered on the source, having an inner radius of 4 pixels and an outer radius of 30 pixels.
From the second orbit on (t$\geq$3\ee{4}~s), when pile-up is no longer present, 
the source count rate was estimated in a circular region with a 30 pixels radius.
The count rate evolution in the later observations (t$>$3\ee{5}~s) 
was obtained using the task {\tt sosta} of the {\tt ximage} package, 
using an extraction region that optimizes the signal to noise ratio of the detection.

All the light curves presented here are background subtracted and corrected for Point Spread Function (PSF) losses, 
vignetting effects and exposure variations.

The 0.3-10 keV light curve is shown in Fig.~\ref{fig:xrtlc} (upper panel). 
WT and PC data were binned to achieve a minimum signal to noise ratio of 8
and 5, respectively. Late points (t$\geq$\e{5}~s) are 3$\sigma$ detections.
We also included the detection taken in Image (IM) mode, 
converting Digital Number (DN) units into count rate \citep{hill06,mangano07}.


\begin{deluxetable}{lcc}
\tabletypesize{\scriptsize} \tablecaption{ Results from X-ray lightcurve fits. 
Two different models have been considered and they are discussed in the text.}
\tablewidth{0pt}
\tablehead{ \colhead{Parameters\ \ \ \ \ \ \ \ } & \colhead{Model 1} & {Model 2}\\
             \colhead{ } & \colhead{Multiple broken PL} & {Simple broken PL} }
\startdata
$\alpha_1$\dotfill 		&  2.44$^{+0.13}_{-0.12}$ 	& 2.45$^{+0.13}_{-0.12}$ \\
t$_{\rm break,1}$ (s)\dotfill 	&  570$\pm$50 			& 730$^{+270}_{-230}$ \\
$\alpha_2$\dotfill 		&  0.09$^{+0.07}_{-0.06}$ 	& 0.72$\pm$0.06\\
t$_{\rm break,2}$ (ks)\dotfill 	&  20.57$^{+0.26}_{-0.11}$ 	& -- \\
$\alpha_3$\dotfill 		&  9.0$^{+2.8}_{-1.0}$ 		& --\\
t$_{\rm break,3}$ (ks)\dotfill 	&  29$\pm$2 			& -- \\
$\alpha_4$\dotfill 		&  0.71$^{+0.08}_{-0.09}$ 	& -- 
\enddata
\label{tab:xlc}
\end{deluxetable}


The light curve has been modeled with power law segments of different slopes,
whose best fit parameters are reported in the second column of  Table~\ref{tab:xlc}.
The best fit model is shown by the solid line in the upper panel of Fig.~\ref{fig:xrtlc}.
We also performed a fit of the light curve with a simple broken power law,
excluding from the analysis the plateau and the following steep drop. 
The model is shown by the dashed line in Fig.~\ref{fig:xrtlc},
with best fit parameters reported in Table~\ref{tab:xlc} (third column).

In both models we used a Gaussian function, centered at t$\sim$530~s and $\sigma\sim$50~s wide,
to model the small bump at the end of the early decay.
The fast rise, after the abrupt drop at $\sim$2\ee{4}~s, 
and the following decay are well described by a Fast-Rise-Exponential-Decay (FRED)
profile, peaking at t$\sim$55~ks.

The hardness ratio light curve is reported in the middle panel of Fig.~\ref{fig:xrtlc}. 
It is defined as the ratio between the source counts in the hard band (H:~1-10~keV) 
and the source counts in the soft band (S:~0.3-1~keV).
An initial hard-to-soft evolution is present followed by a hardening of the spectrum,
which corresponds to the small bump in the X-ray light curve at $\sim$530~s.
The hardness ratio shows an increasing trend at the beginning of the observed plateau
and spectral variations throughout it. The very steep drop displays a hard-to-soft evolution.
Later points (t$\gtrsim$\e{5}~s) have a harder spectrum, maybe due to a late flaring activity, 
as the fluctuations in the light curve suggests.
At times later than 2\ee{5}~s the afterglow evolution does not show any further spectral variation.


\begin{deluxetable}{cccccc}
\tabletypesize{\scriptsize} \tablecaption{ Results from X-ray spectral fits\tablenotemark{a}.}
\tablewidth{0pt}
\tablehead{ \colhead{X-ray} & \colhead{Instr.} & \colhead{Start} &
            \colhead{Stop} & \colhead{$\Gamma_{\rm X}$} & \colhead{\chisq/dof}\\
            \colhead{Phase} & \colhead{Mode} & \colhead{(s)} & 
	    \colhead{(s)} & \colhead{ }  & \colhead{ } }
\startdata
Early & WT &   99.7 & 117 & 1.84$\pm$0.11 & 14/21 \\
decay & WT &  117   & 137 & 1.92$\pm$0.14 & 23/17 \\
      & WT &  137   & 177 & 2.16$\pm$0.14 & 21/18 \\
      & WT &  177   & 265 & 2.21$\pm$0.14 & 6/14 \\
      & PC &  266   & 644 & 1.7$\pm$0.3 & 7/6 \\

\hline

Plateau & PC &  4043 &  5237  & 2.32$\pm$0.17 & 12/15  \\
        & PC &  5237 &  6426  & 2.05$\pm$0.16 & 21/17  \\
        & PC &  9825 &  10997 & 2.06$\pm$0.14 & 23/16  \\
        & PC & 10997 &  12210 & 1.92$\pm$0.13 & 13/16 \\
        & PC & 15607 &  16896 & 2.16$\pm$0.14 & 28/19  \\
        & PC & 16896  &  17993 & 2.17$\pm$0.14 & 20/17  \\
\hline
Drop    & PC & 21394 & 29559 & 2.12$\pm$0.15 & 14/15 \\
\hline 
Shallow & PC & 5.8\ee{5} & 1.4\ee{6} & 2.17$\pm$0.20 &  -- \\
decay\tablenotemark{b}   &    &   &   &  & 
\tablenotetext{a}{Spectra were modelled with an absorbed power law. Galactic and intrinsic absorption
were kept fixed at the value of 1.86\ee{20}\cm{-2} and 2.6\ee{21}\cm{-2}, respectively.}   
\tablenotetext{b}{The Cash statistic was applied.}
\enddata
\label{tab:pha}
\end{deluxetable}


\subsubsection{Spectral analysis}\label{sec:pha}

In order to quantify the spectral variations seen in the hardness ratio light curve,
we performed a time-resolved spectral analysis.

Source and background spectra were extracted from the same regions 
used to create light curves (\S~\ref{sec:lc}). 
The relevant ancillary response files were generated using the task {\tt xrtmkarf}.
Time intervals were selected according to light curve phases
and to have at least 400 net counts each. Only the first PC orbit 
spectrum (from 266~s to 644~s) has a lower statistical quality ( $\sim$200 source counts), 
due to the presence of pile-up.
Spectral channels were grouped so to have at least 20 counts each.
The \chisq~statistic was applied.

All the X-ray spectra can be modeled with an absorbed power law. 
The Galactic absorption component was kept fixed at the value 
of 1.86\ee{20}\cm{-2} \citep{nh90}; an additional redshifted absorption component,
modeling the host intrinsic absorption, was also included.
In order to estimate the host absorption 
we extracted a WT (from 100~s to 265~s) and a PC spectrum (from 4~ks to 30~ks).
Since at low energies ($\sim$0.5~keV) XRT spectra may be affected by 
calibration uncertainties\footnote[1]{http://heasarc.gsfc.nasa.gov/docs/swift/analysis/},
we performed our analysis excluding energy channels in the range 0.45-0.55~keV.
From the joint spectral fit we obtained an intrinsic absorption of 
N$_{\rm H}^{\rm host}$=(2.6$\pm$1.1)\ee{21}\,\cm{-2}. 
This value was derived assuming a host galaxy with solar metallicity.

We then performed time-resolved spectral fits keeping the absorption
fixed as above and leaving only the photon index
$\Gamma_{\rm X}$ and the normalization as free parameters.
We also tested whether the initial softening of the spectrum could be 
due to a decreasing intrinsic absorption, as previously witnessed in other GRBs 
(Cusumano et al. 2007, Campana et al. 2006),
but the low number of counts did not allow us to constrain the N$_{\rm H}^{\rm host}$ behavior.

In order to estimate the spectral index during the final shallow decay,
we extracted a spectrum in the time interval from 5.8\ee{5}~s to 1.4\ee{6}~s,
using a circular region of 5 pixels radius centered on the source position. 
Because of the low number of counts ($\sim$120) 
and the negligible background contamination, the Cash statistic was applied \citep{cash79}.
We obtained $\Gamma_{\rm X}$=2.17$\pm$0.20.

The selected time intervals and results from the spectral fit 
are listed in Table~\ref{tab:pha}.
The photon index variations with time are reported 
in the lower panel of Fig.~\ref{fig:xrtlc}.



\begin{deluxetable}{ccccc}
\tabletypesize{\scriptsize} \tablecaption{\bf \swift~UVOT photometry of GRB~070110 afterglow.}
\tablewidth{0pt}
\tablehead{ \colhead{Start} & \colhead{$\Delta$t} & \colhead{Magnitude\tablenotemark{a}} 
            & \colhead{Error\tablenotemark{b}} & \colhead{Filter} \\
            \colhead{(s)} & \colhead{(s)} & \colhead{ } & \colhead{}  & \colhead{} }
\startdata
103	&	100	&	19.90	&	0.31	&	White	\\
4249	&	200	&	19.18	&	0.26	&	White	\\
5681	&	200	&	19.69	&	0.18	&	White	\\
\hline									
85	&	9	&	17.35	&	UL	&	V	\\
208	&	400	&	20.28	&	0.45	&	V	\\
4658	&	200	&	18.94	&	0.33	&	V	\\
6091	&	200	&	18.83	&	UL	&	V	\\
22298	&	18665	&	20.38	&	0.27	&	V	\\
161578	&	47181	&	21.52	&	0.32	&	V	\\
\hline									
4044	&	1632	&	20.33	&	0.30	&	B	\\
11644	&	598	&	20.59	&	0.21	&	B	\\
28983	&	603	&	21.32	&	0.36	&	B	\\
114817	&	12775	&	22.08	&	0.32	&	B	\\
\hline									
662	&	15	&	19.23	&	0.51	&	U	\\
5272	&	200	&	19.63	&	0.22	&	U	\\
10733	&	906	&	19.84	&	0.14	&	U	\\
17423	&	602	&	20.22	&	0.23	&	U	\\
28072	&	906	&	20.42	&	0.22	&	U	\\
34762	&	603	&	20.84	&	0.31	&	U	\\
235449	&	135190	&	22.58	&	UL\tablenotemark{c}	&	U	\\
495668	&	81911	&	22.07	&	UL	&	U	\\
\enddata
\tablenotetext{a}{Values are not corrected for extinction.}
\tablenotetext{b}{Errors at the 68\% confidence level or 
                  3$\sigma$ upper limits (ULs) are given.}
\tablenotetext{c}{Our re-analysis using the refined
                  optical position gives a 3$\sigma$ UL instead
		  of the detection quoted in \citet{report}.}	  
\label{tab:uvot}
\end{deluxetable}



\begin{figure*}
\centering
\includegraphics[angle=270,scale=0.65]{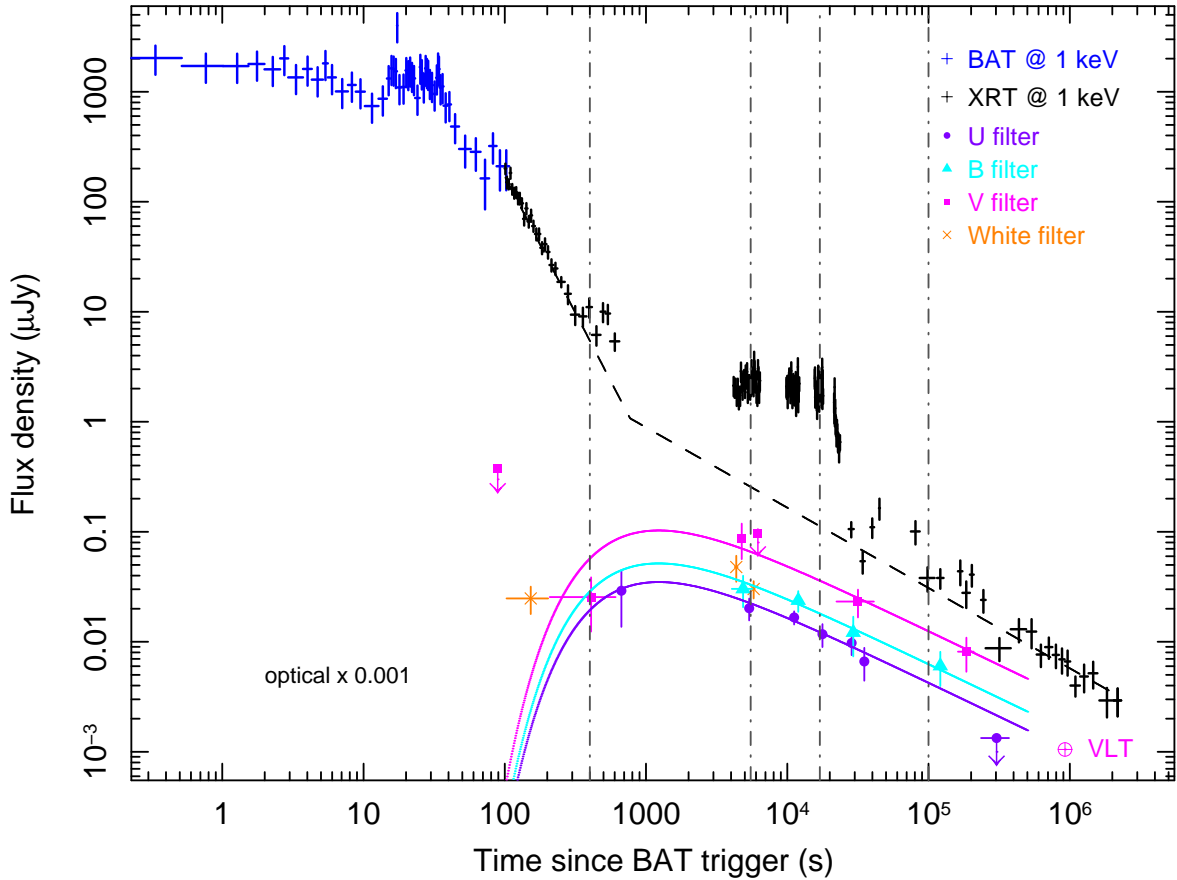}
\caption{Combined BAT, XRT and UVOT light curves.
         BAT and XRT count rates were converted into monochromatic fluxes 
	 using the results from the spectral analysis. 
	 UVOT magnitudes were corrected for the Galactic extinction
	 and converted into flux densities at the central wavelength of each filter.
         In the plot UVOT flux values have been scaled by a factor \e{-3}.
	 The optical best fit models, discussed in the text, are shown by solid lines.
	 The dashed line shows the broken power law model (model 2).
	 The vertical dashed-dotted lines mark the times at which SEDs were computed.
	 }
\label{fig:swiftlc}
\end{figure*}


\subsection{Ultraviolet/Optical data}\label{sec:uvot}

Independent measurements of the position of the afterglow were
made after summing the images taken during the first day
after the trigger for the White, $V$, $B$, and $U$ filters.
The mean position is \RA{00}{03}{39.23}, \decl{-52}{58}{26.9} with an
estimated 1$\sigma$ uncertainty of ~0.25'' in each direction,
based on the rms deviation of the four measurements.
This position is consistent with the ground-based position given in \citet{gcn6015}.

We performed photometry 
using a circular aperture of radius  $2.5''$ centered on the position of the optical afterglow.
This choice allows us to minimize the background contribution.
Aperture corrections were computed using stars in the images to convert 
the $2.5''$ photometry to a $6''$ aperture for the $V$, $B$ and $U$ filters, 
and a $12''$ aperture for the White filter.
Adjacent exposures with $\Delta$T/T$<$0.1 were coadded 
to improve the signal to noise.

The instrumental magnitudes were transformed to Vega magnitudes using the standard
photometric zero points in the \swift/UVOT calibration 
database\footnote[2]{http://heasarc.gsfc.nasa.gov/docs/heasarc/caldb/swift/docs/uvot/index.html}.
Table~\ref{tab:uvot} contains the results of the UVOT photometry.
The reported errors and upper limits (ULs) are at 1$\sigma$ and 3$\sigma$ confidence level, 
respectively.

The Galactic reddening in the burst direction is E(B-V) = 0.014 \citep{schlegel98}.
Using the value from the extinction curve in \citet{pei92} for the Milky Way
at the central wavelength of each filter, we estimated the extinction for the UVOT filters to be
A$_V$ = 0.04~mag, A$_B$ = 0.06~mag and A$_U$ = 0.07~mag.

Fig.~\ref{fig:swiftlc} presents the multi-wavelength light curve of GRB~070110,
including $\gamma$-ray, X-ray and UV/optical data.
Both BAT and XRT light curves were converted into flux units using 
the best fit spectral models. The $V$, $B$ and $U$ magnitudes were corrected 
for Galactic extinction along the line of sight and then converted to 
monochromatic fluxes at the central wavelength of each filter. 
Since the afterglow is not detected blueward of the $U$ filter, we took
4450\AA\ as the effective mid-wavelength of the White filter. 
The conversion factor from count rate to monochromatic flux density
depends on the afterglow spectral shape and the White filter effective area curve
and it was found to be 0.026~mJy~s~cts$^{-1}$.

With the exception of the first detection in the $U$, $V$ and White filter,
the UVOT light curves show a decaying behavior that can be well described by a simple power law.
An independent fit of the three light curves gives the following temporal slopes:
$\alpha_V$=0.65$\pm$0.14, $\alpha_B$=0.53$\pm$0.13, $\alpha_U$=0.54$\pm$0.15.
We note that the UVOT late slopes are consistent, within the errors, 
to the late-time X-ray slope in Table~\ref{tab:xlc}.
Including in the fit also the late VLT detection in the $V$ band, reported in \citet{gcn6021}, 
the slope $\alpha_V$ steepens to 0.88$\pm$0.09. 
Alternatively, we attempted a fit with a broken power law, which gives
an initial decay of 0.62$\pm$0.16, steepening at $\sim$150~ks to 1.3$\pm$0.2.
Both models provide a good description of the light curve in the $V$ band.

A joint fit of the three UVOT light curves (U, B, and V) has been performed
using the functional form of \citet{willi07}.
We allowed the normalizations to vary independently, tying other parameters.
Upper limits and the VLT detection were not considered in the fit.
 We found that our light curves follow a power law decay 
with a common slope of 0.63$\pm$0.13.

The best fit models of the three UVOT light curves (U, B, and V)
with the \citet{willi07} function are shown by solid lines in Fig.~\ref{fig:swiftlc}.
The fast rise of the optical emission is poorly sampled and we note
that the first detection in the White filter suggest a more gentle 
onset of the afterglow emission.

For a comparison we report the broken power law model (model 2),
described in \S~\ref{sec:lc} (dashed line in Fig.~\ref{fig:swiftlc}).


\begin{figure*}
\centering
\includegraphics[scale=0.4]{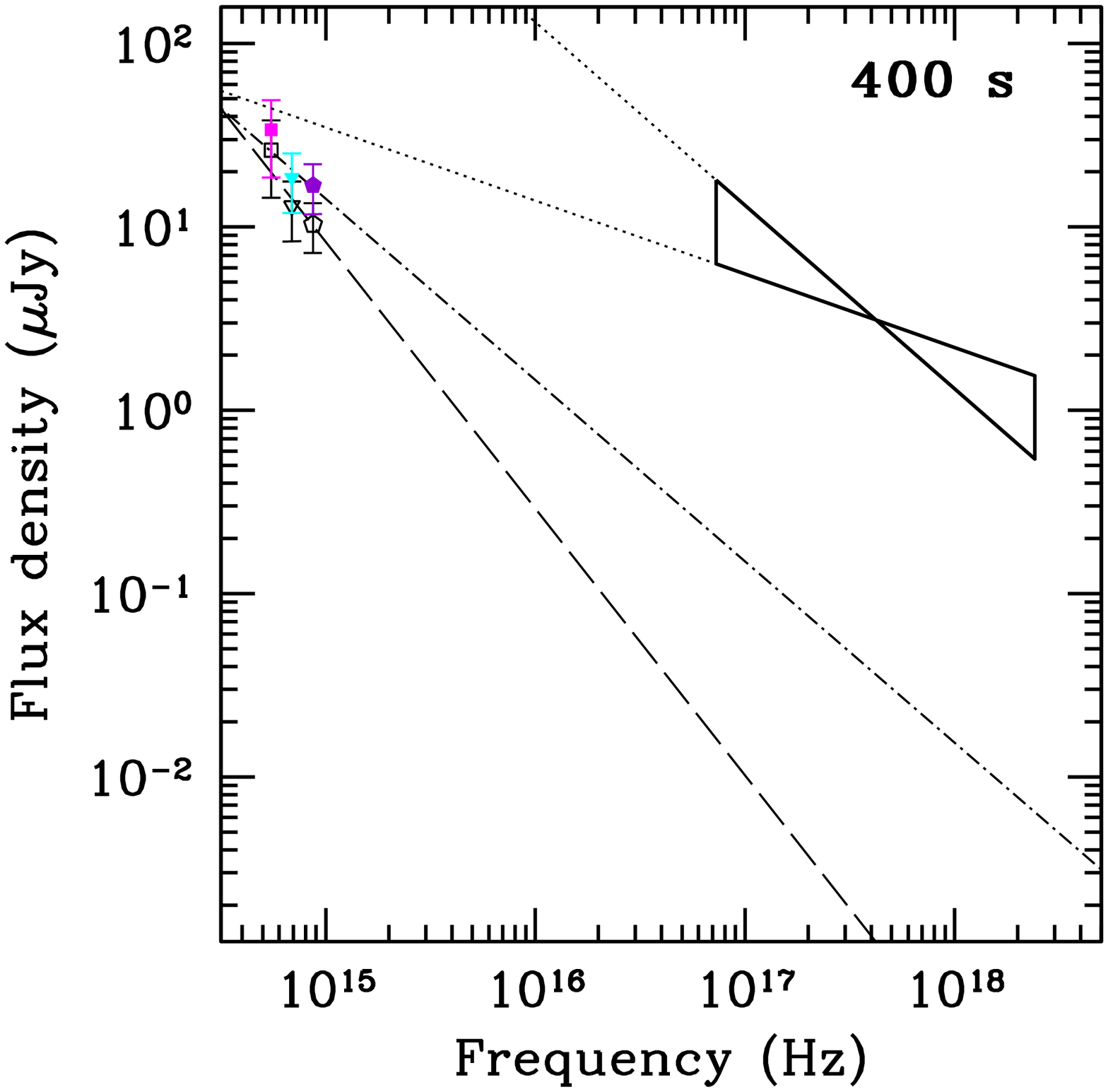}
\includegraphics[scale=0.4]{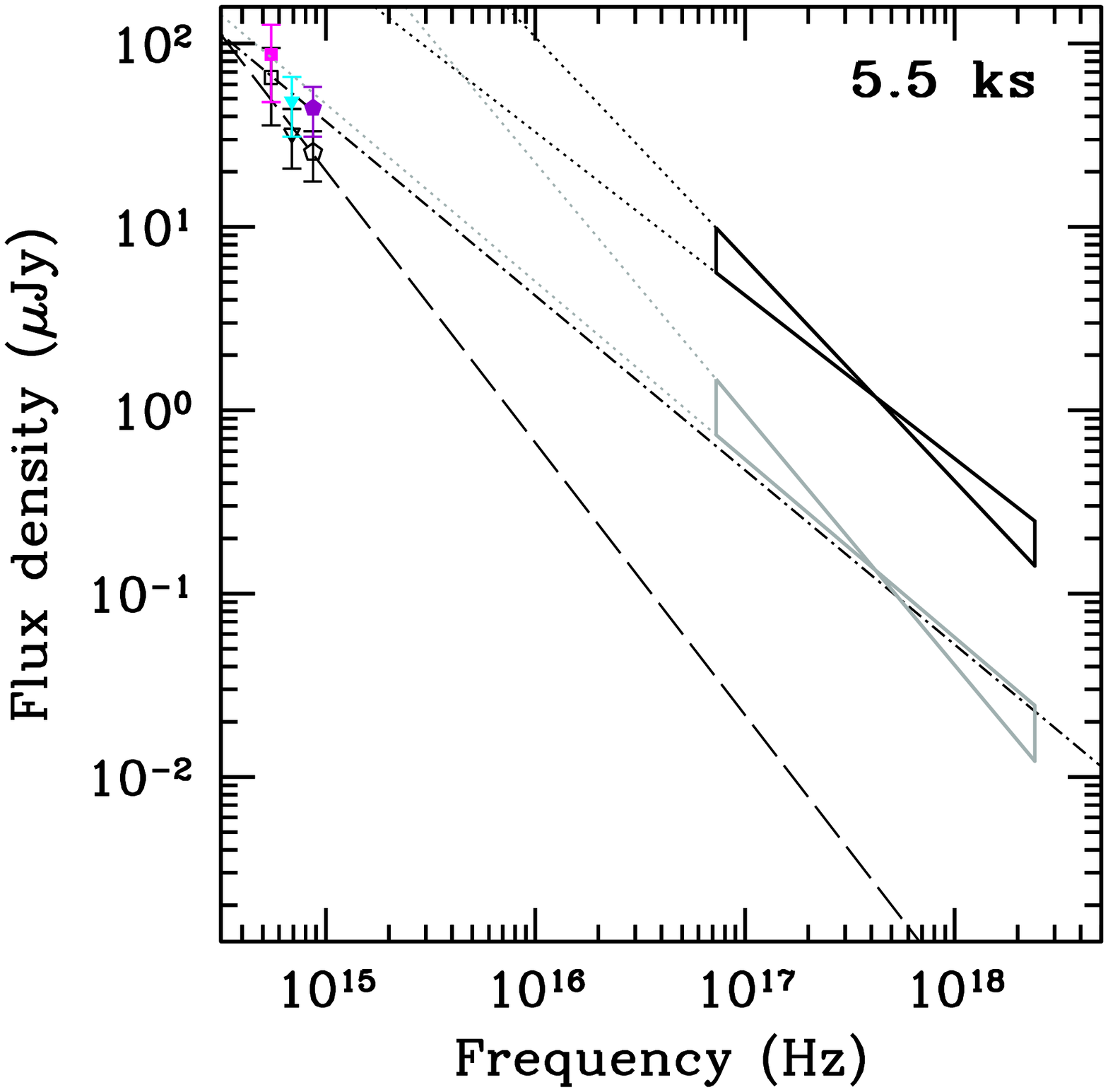} 
\includegraphics[scale=0.4]{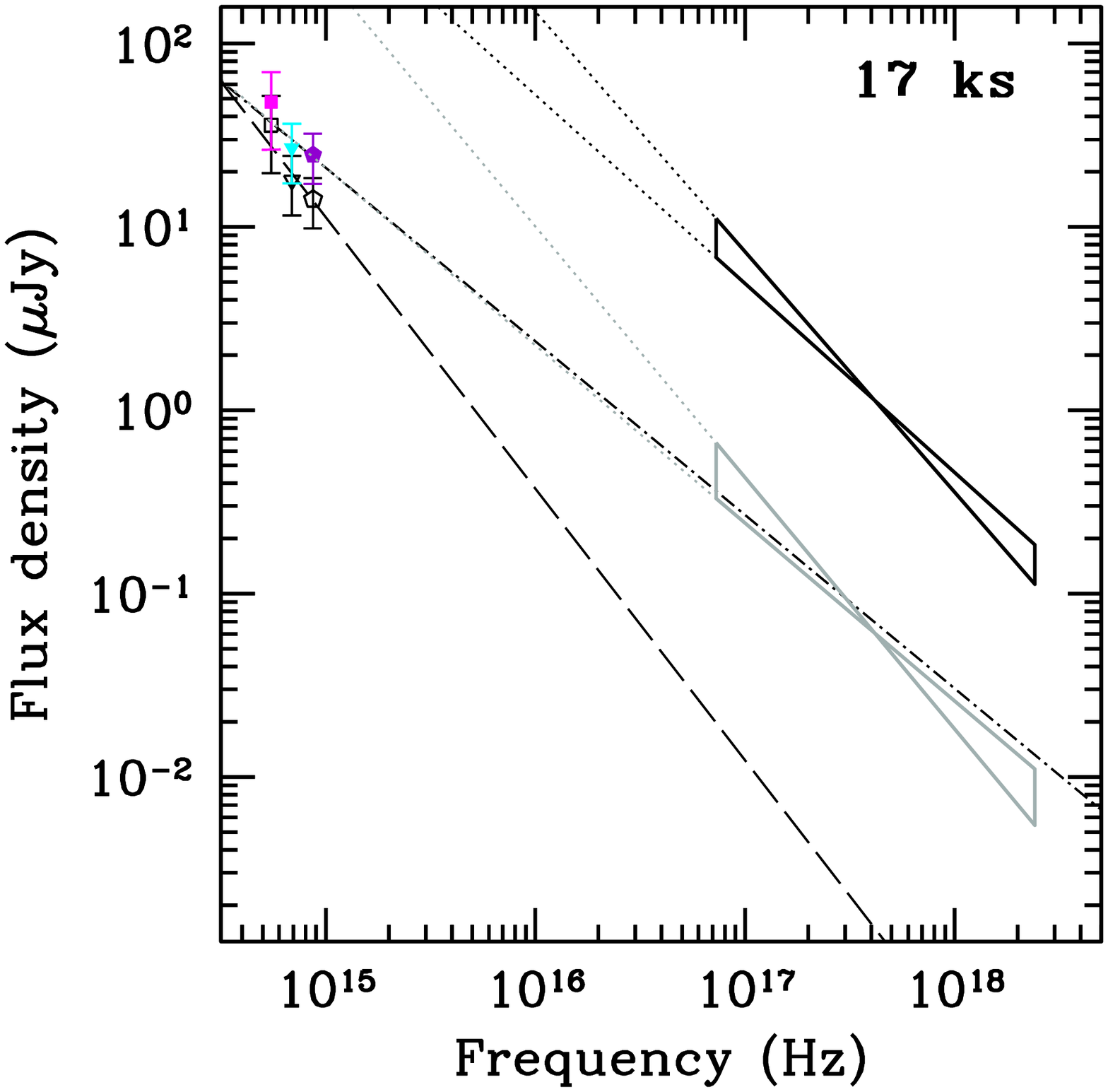}
\includegraphics[scale=0.4]{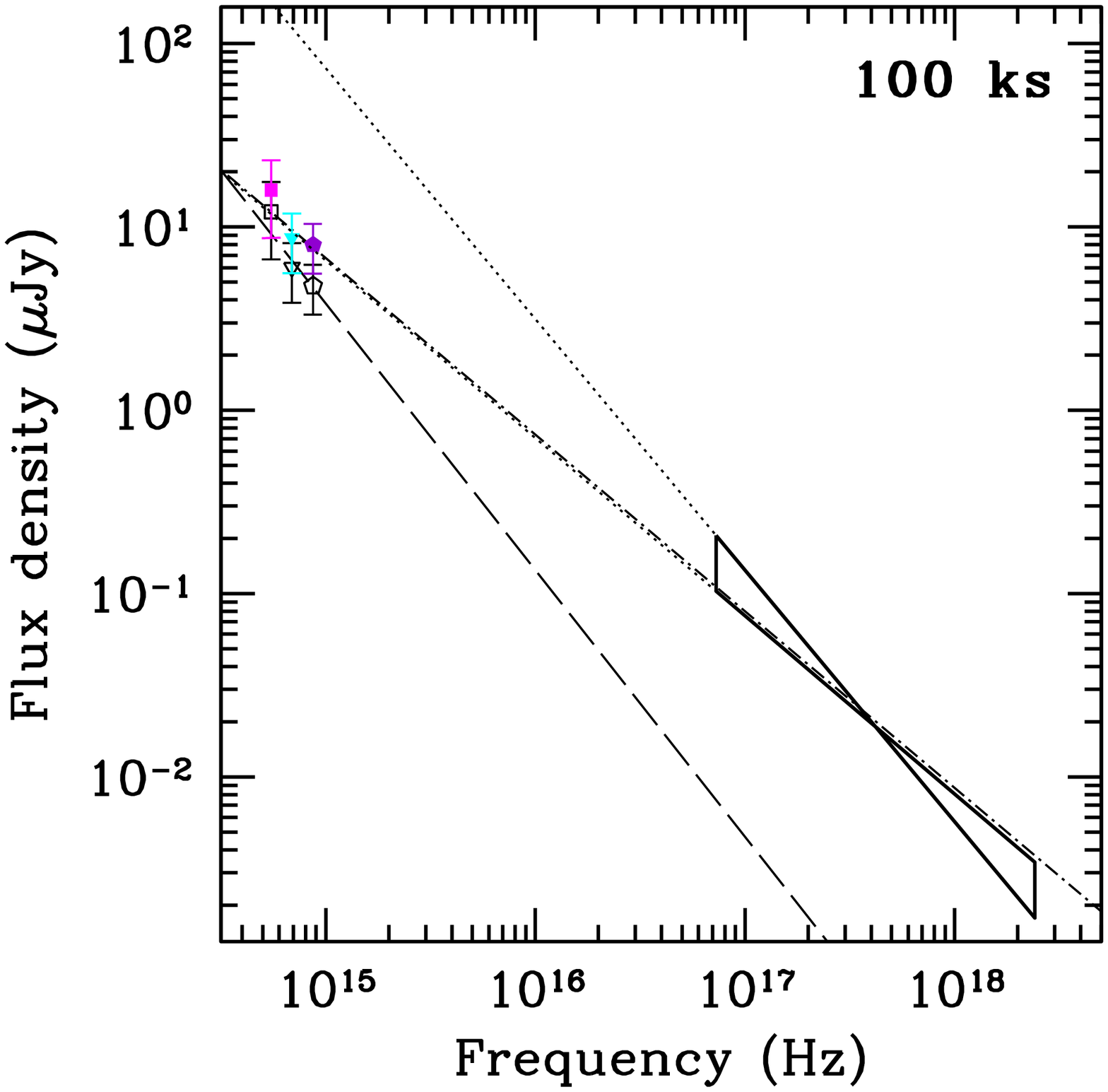}
\caption{
Optical, ultraviolet and X-ray SEDs at 400 s, 5.5 ks, 17 ks, and 100 ks.
UVOT points have been corrected for the Galactic extinction A$_V$=0.04~mag
(empty symbols) and for the host extinction A$_V$=0.08~mag (filled symbols) along the line of sight .
The dashed line is the best fit of the optical data corrected only for Galactic extinction.
The dot-dashed line is the best fit of the optical data corrected for both the
Galactic and the host intrinsic extinctions.
The solid black lines define the cone corresponding to the 90\% uncertainty on the 
spectral slope in the X-ray band. This cone is extrapolated to the optical 
band through the dotted lines. 
The X-ray fluxes, extrapolated at t=5.5~ks and t=17~ks according to our model 2,
and the corresponding error cones are also shown (grey solid lines).
}
\label{fig:sed}
\end{figure*}


\subsection{Optical/X-ray Spectral Energy Distributions}\label{sec:sed}

We calculated the Spectral Energy Distribution (SED) at four 
representative times: t=400~s during the initial decay, 
t=5.5~ks and t=17~ks at the beginning and at the end of the plateau phase,
and t=100~ks during the shallower decay.
The four times are marked by vertical lines in Fig.~\ref{fig:swiftlc}.

Fig.~\ref{fig:sed} shows the four SEDs derived from the optical and X-ray light curves.
Flux values in the UVOT filters at the selected times are corrected 
for Galactic extinction (empty symbols) and
were derived from the best fit models shown in Fig.~\ref{fig:swiftlc}.
At a redshift of $z$=2.352 the Ly$\alpha$ forest is present throughout
the $U$ filter. Therefore, in constructing the SEDs we have corrected
the $U$ band count rate for the expected intergalactic Ly$\alpha$
transmission of 86\% over the $U$ passband \citep{madau95}.

X-ray data were converted into flux units for each time selecting the appropriate 
photon index value from Table~\ref{tab:pha}. 
X-ray fluxes and their error regions are shown by the black cones.
The contribution of the host galaxy to extinction have been estimated as 
the additional extinction required to have optical and X-ray data lying
on the same power law at t=100 ks. 
Accounting for a host extinction of $A_V$=0.08 mag with the extinction curve for the
Small Magellanic Cloud (SMC; \citealt{pei92}) provides a better match 
between optical and X-ray data. The entire data set at the later times 
seems consistent with a common physical origin.
UVOT fluxes corrected for both the Galactic and the intrinsic extinctions 
are shown by filled symbols. 
The best fit power law of the optical SED is shown by the dot-dashed line.

At early times (t=400~s) the optical data are not consistent with the extrapolation 
of the X-ray spectrum to low energies.
At 5.5 ks and 17 ks, during the apparent plateau, the optical and X-ray spectral distributions are also completely 
inconsistent with one another, implying different origins for the optical and X-ray photons.

In \S~\ref{sec:lc} we have shown that the X-ray light curve, excluding the plateau, 
can be fitted with a simple two component broken power law plus flares. 
Since the slope of this power law is marginally consistent with that observed in the optical bands, 
it follows that the SED consistency observed at late times also applies to this component. 
This is illustrated in Fig.~\ref{fig:sed} (top right panel and bottom left panel), 
where we have extrapolated the late-time SED to 5.5 and 17 ks, 
using light curve model 2 (third column in Table~\ref{tab:xlc}) to evaluate the X-ray count rates.
These values were derived using the photon index value at late times (t$>$580 ks).
The extrapolated values and their error regions are shown by the grey cones.

If we assume for the host galaxy of GRB~070110 the same gas-to-dust ratio 
measured in the SMC, N$_{\rm H}/A_V$=1.6\ee{22}~cm$^{-2}$~mag$^{-1}$ \citep{smc00},
an intrinsic hydrogen column density of 1.3\ee{21}~cm$^{-2}$ is inferred from
the estimated rest frame visual extinction A$_V$=0.08 mag.

The intrinsic absorption derived from fitting the X-ray spectra is instead (1.6$\pm$0.7)\ee{22}~cm$^{-2}$
for a SMC-like absorber, higher than the one expected from the gas-to-dust relationship.
We derived the latter absorption value modeling the intrinsic X-ray absorption 
spectral component with the model {\tt tbvarabs} in XSPEC \citep{xspec}, 
keeping the absorber abundances fixed to SMC values ($\sim$Z$_{\odot}$/8).

The high X-ray absorption column and low optical extinction measured for GRB~070110 
require a higher than expected gas-to-dust ratio, and are indicative of an interstellar medium
significantly different from the one observed in the local universe.
Such difference could be ascribed to the action of the GRB radiation on its immediate environment
\citep{wax00,perna03} or, alternatively, to an intrinsically higher gas-to-dust ratio 
in GRBs birth sites \citep{watson06}.
Discrepancies between optical and X-ray extinctions 
in GRBs environments were first found by \citet{galama01} and then confirmed 
by following studies on large GRBs samples (e.g. \citealt{kann06,starling07,taglia07}).

\section{Discussion}

\subsection{The Early X-Ray Tail}

As shown in Fig.~\ref{fig:xrtlc} (upper panel), the X-ray light curve is composed of four distinct
components. It starts with (I) an early steep decay component with
$\alpha=2.44\pm0.13$ following the prompt gamma-rays. 
The early decay is followed by (II) a plateau with an essentially constant flux which
extends up to t$\sim$20 ks. Luminosity fluctuations are seen throughout the plateau phase. 
Following the plateau is (III) a remarkable steep drop with a temporal index of $\sim$9.0. 
After this steep fall the afterglow light curve rises again, showing a late flare,
then enters (IV) a more normal decay segment with $\alpha\sim$0.7.

The hardness ratio light curve (Fig.~\ref{fig:xrtlc}, middle panel)
shows an initial hard-to-soft spectral evolution during the early decay.
Our spectral analysis confirms a softening of the spectral index $\beta_{\rm X}$
from a value of 0.8 to 1.2.  
The early X-ray decay has been generally interpreted as the prompt emission tail,
due to the delay of propagation of photons from high latitudes with respect to the line of sight
\citep{taglia05,nousek06,obrien06,zhang06}.

The curvature effect alone, which suggests $\alpha$=2+$\beta$ \citep{kumar00}, 
cannot explain both the spectral evolution feature and the shallower than expected
temporal slope.
Such behavior has been seen in other GRBs detected by \swift\ 
and, as reported by \citet{tail06}, 
spectral evolution seems a common feature in early, bright GRB tails
(e.g. GRB~060614, \citealt{mangano07b}).
Phenomenological models involving an underlying central engine afterglow 
with a steep spectral index or an internal shock afterglow with a 
cooling frequency passing through the XRT window
may be candidates to interpret these spectral evolutions \citep{tail06}.

\subsection{The Plateau Phase}

The observed plateau is a feature of great interest. 
This component displays an apparently constant intensity extending up to t$\sim$20~ks, 
followed by an abrupt drop with a very steep decay index. 
Compared with the canonical XRT light curves GRBs observed by \swift\ 
\citep{nousek06,obrien06,zhang06,willi07},
such a steep decay following a plateau is unique. 

In most other Swift GRBs, the plateau is followed by a ``normal'' decay 
which could be generally interpreted as the standard external forward shock afterglow. 
The plateau in those cases is therefore consistent with a refreshed shock
\citep{remes98,zhang06,nousek06}
as the total energy in the blast-wave is replenished with time. 

The abrupt steep decay in GRB~070110, on the 
other hand, is inconsistent with an external shock origin of the plateau. 
This is because, within the external shock model, it is impossible
to produce a decay slope as steep that observed.
The steepest possible decay from an external shock occurs
after a jet break.
The post jet break decay slope can be as steep as $\alpha \sim p$, where $p$ is the 
electron spectral index \citep{rhoads99,sari99} and afterglow
modeling suggests that $p$ is typically between 1.5-3. 

A steep decay may be expected when the blast-wave leaves a density clump and falls 
into a very tenuous medium. In such a case, the curvature effect
\citep{kumar00} defines the decay slope. 
However, detailed numerical calculations suggest that the zero-time point of the
external shock scenarios cannot substantially be shifted from the GRB trigger time 
\citep{zhang06,lazzati06,kobayashi07}.
As a result, the expected steep decay index within
this scenario is $\alpha$$\sim$2+$\beta$$\sim$3, far shallower than the
observed value.

By assuming that the steep decay following the plateau is due to the curvature effect,
we performed the same test of \citet{liang06}, who tested the
internal origin of X-ray flares.
We found that to satisfy the $\alpha=2+\beta$ condition 
the zero time t$_0$ needs to lie in the range 16-18~ks after the trigger, 
slightly before the beginning of the abrupt drop.
The very steep decay component therefore strongly suggests 
that the plateau is of {\em internal} origin, since t$_0$
is allowed to shift to much later epochs within the internal 
scenarios \citep{zhang06,liang06}.


\begin{figure}
\centering
\includegraphics[angle=270,scale=0.35]{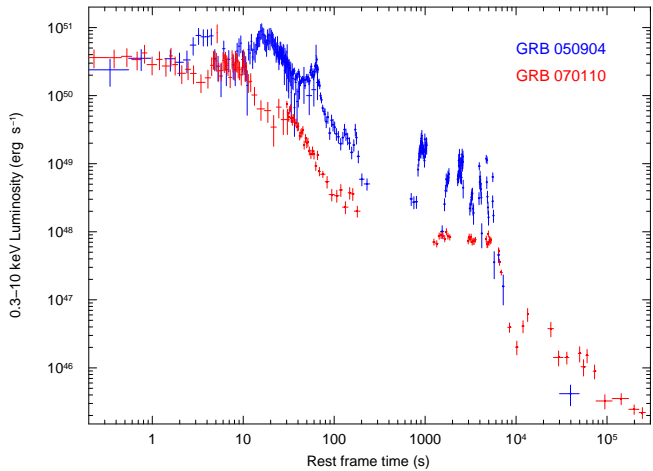}
\caption{Combined BAT and XRT light curves of GRB~050904 ($z$=6.29) and GRB~070110 ($z$=2.35) in the source rest frame.}
\label{fig:compare}
\end{figure}


Several other GRBs have previously shown similar, but not exactly
the same, behavior.  
The afterglow of GRB~051117A \citep{goad07}
shows a sharp drop of the X-ray flux at $\sim$\e{4}~s, 
followed by a shallower decay ($\alpha$$\sim$0.7).
However, in this case the early X-ray light curve seems dominated by flaring activity
and it does not show any evident plateau. 
Fig.~\ref{fig:compare} compares the rest frame light curves of both GRB~050904 and GRB~070110.
GRB~050904 \citep{cusumano06,cusumano07} shows a remarkably similar behavior to GRB 070110, with a 
rapid decay following the multiple episodic X-ray flares.  
However, GRB~070110 displays a much smoother emission episode before the rapid decay. 
While the steep decay following the flares in  GRB~050904 is expected, 
the dropoff after the plateau in GRB 070110 is not.

On the basis of \swift~data, collected in portions lasting less than an hour,
we cannot definitely rule out that the observed plateau
is due to the same mechanism producing erratic X-ray flares 
in other GRBs, such as GRB~050904. 
The hardness ratio (Fig.~\ref{fig:xrtlc}, middle panel)
appears to rise slowly at the start of the plateau ($\sim$4~ks)
and a decreasing trend is visible during the sharp decay ($\sim$20~ks).
Such behavior is similar to the one seen in the hardness ratio
light curves of X-ray flares, although timescales associated with a
single flare are usually shorter.
Attributing the whole emission between 4~ks and 20~ks 
to a superposition of many flares allows
the final drop to be explained with the curvature effect, resetting the
reference time t$_0$ just before the rapid decay.
In this case, the smoothness of the flat phase could be a selection effect,
due to lacking full coverage of the light curve.
The hypothesis of flares blending to form an apparent plateau seems however very unlikely,
requiring a high degree of homogeneity between successive X-ray flares,
since no large amplitude variations have been observed (see Fig.~\ref{fig:xrtlc}).

We conclude that GRB~070110 robustly suggests new properties of the GRB central engine: 
it must be able to last for an extended period of time 
(in this case up to 8\ee{3}~s in the source rest frame) 
with an essentially constant radiation power.

Previous \swift~observations of X-ray flares \citep{burrows05, barthelmy05,
falcone06,romano06,campana06,cusumano06,cusumano07} have revealed that the GRB central engine
can indeed be active to such a late epoch. 
However, the energy output (at least the observed luminosity) 
in previous cases has to be intermittent. 
The mechanisms to interpret these intermittent X-ray flares therefore
require that the central engine accretes episodically, either due to
fragmentation or density fluctuation of the accretion disc 
(e.g. \citealt{king05,perna06,rosswog06}), or due to modulation of the magnetic fields 
near the accretor \citep{proga06}. Alternatively, the intermittent
behavior of the central engine may be related to the unsteady magnetic
activity of the central engine (e.g. \citealt{dai06}). 
None of these suggestions seems to be able to interpret the internal plateau
in GRB~070110 straightforwardly. Rather we require that the central engine is
continuously active for a long period of time. 

Such long-term central engine activity has been discussed before 
(e.g. \citealt{zhang01} for a general discussion of a central engine with decaying
luminosity L(t)$\propto$t$^{-q}$). The most straightforward example
is a spinning-down pulsar \citep{dai98,zhang01}.
This model has been introduced to interpret the external plateau as due to refreshed shocks, 
although those data can be also interpreted without introducing such a long-term central engine
\citep{remes98,granot06}. The discovery of
the internal shock origin of the plateau in GRB~070110 breaks the degeneracy of the refreshed shock
interpretation,
and gives the first direct evidence of a continuous long-lasting central engine. 

We propose that the engine powering the plateau could be a spinning-down pulsar,
which has a constant luminosity lasting for an extended period of time 
\citep{shakira,zhang01}.
The duration of the plateau depends on the unknown pulsar parameters, but given a 
reasonable radiation efficiency, the luminosity ($\sim$\e{48}~erg\,s$^{-1}$)
and the observed duration ($\sim$16~ks)
of the plateau are consistent with the parameters of a new-born
magnetized millisecond pulsar as the GRB central engine.

According to \citet{zhang01}, 
the continuous injection luminosity $L_{{\rm em},0}$ and 
the charateristic time scale ${\cal T}_{\rm em}$, when the plateau breaks down,
are related to the pulsar initial parameters:
\begin{equation}
L_{{\rm em},0}\simeq 10^{49}~B_{{\rm p},15}^2P_{0,-3}^{-4}R_6^6~{\rm erg~s^{-1}}.
\label{Leem}
\end{equation}
\begin{equation}
{\cal T}_{\rm em}\simeq2.05\times10^3~ I_{45}B_{{\rm p},15}^{-2}P_{0,-3}^2R_6^{-6}~{\rm s}.
\label{Teem}
\end{equation}
where $B_{\rm p}=B_{{\rm p},15}\times10^{15}$~G is the dipolar field strength at the poles,
 $P_{0,-3}$ is the initial rotation period in milliseconds,  
 $I_{45}$ is the moment of inertia in units of \e{45}~g\,cm$^2$, and $R_6$ is stellar radius in units of \e{6}~cm.

If we assume that a significant fraction of the spin-down luminosity is emitted in the X-ray band
and take standard values of $I_{45}\sim1$ and $R_6\sim1$,
from the plateau parameters we can infer a pulsar initial period of $P_0\lesssim$1~ms
and a magnetic field $B_p\gtrsim3\times10^{14}$~G. 
The energy of the plateau puts only a lower limit to the real energy
of the central pulsar $L_{{\rm em},0}$  and without a good estimate of the radiation efficiency
it is not possible to constrain better the pulsar parameters.
It is worth to note that our estimate of $P_0$ is very close to the break-up period for a neutron star,
$P_{\min}=$0.96~ms \citep{lattimer04}, 
so the derived pulsar parameters can be taken as good approximations of the right values.

The observed decay slope following the plateau is much steeper than the model prediction ($\alpha$$\sim$2).
However, accounting for detailed energy dissipation mechanisms and 
possible magnetic field decay at the central engine can steepen
the decay (Zhang et al., in preparation).
The energy dissipation mechanism of a spinning-down
pulsar before deceleration is not well studied, but it may be related
to the breakdown of the magneto-hydrodynamic condition in the magnetized outflow
\citep{usov94,spruit01,zhang02b} and 
powered by magnetic reconnections \citep{dren02}. 
Instability inside the energy dissipation regions would induce 
fluctuations in the emission region, giving rise to the flickering
light curve and the hardness ratio variations on the plateau.

\subsection{The Final Shallow Decay}

The late X-ray light curve shows a shallow power law decay,
on which a flare at $\sim$55~ks and mini-flares up to $\sim$200~ks are superimposed. 
These flares could be again interpreted as late central engine activities
\citep{burrows05,romano06,falcone06,zhang06,fan05}
which are likely related to the episodic accretion processes 
and may be different from the spindown-powered plateau.
 
From the SEDs analysis (\S~\ref{sec:sed}) we derived that in the time interval 5-20 ks we observe 
an internal afterglow component, which contributes strongly to the X-ray but not the optical band, 
and a second afterglow component, likely of external shock origin, which tracks the optical light curves. 
This component dominates after the late steep drop, when the emission from the central engine switched off.
 The late optical/X-ray spectrum (t=100~ks) can be described by a continuous power law
($\beta$=1.00$\pm$0.14),
indicating that the optical and the late X-ray afterglow may arise from the same physical component.

The shallow temporal slopes in the X-ray band ($\alpha_{\rm X}\sim$0.7) 
and in the optical bands ($\alpha_{\rm opt}$=0.63$\pm$0.13) are consistent.
However we note that the observed temporal decays are not in agreement with the standard closure relations 
(e.g. \citealt{zhang04}), which predict a temporal slope $\alpha\gtrsim$1 for both the X-ray and the optical light curves. 
Such shallow behavior at late times is seen in $\sim$50\% of the X-ray light curves
observed by \swift, as noted by \citet{willi07}, and it may suggest that the external shock
is kept refreshed by a long-lasting energy injection.


\section{Conclusion}

GRB~070110 is a long burst (T$_{90}$$\sim$90~s) with a standard behavior
in the $\gamma$-ray band,
that shows extraordinary properties of the X-ray afterglow.
The presence of a plateau followed by an unexpected steep decay 
implies a new emission component from a long-lasting central engine.
Such constant emission may be generated by a spinning down pulsar.
This mechanism indeed allows us a better interpretation
of the GRB070110 phenomenology than the superposition of multiple flares.
However the naive picture we discussed cannot account for the rapid decay 
at the end of the plateau, indicating that a more detailed 
theoretical study is needed. 
A drop of the radiation efficiency, caused by the pulsar deceleration,
or a decaying magnetic field of the new-born system 
could help to explain the steep decay rate.\\


\acknowledgements{}

We wish to thank the anonymous referee for his/her careful reading
of the paper. We also thank Daniele Malesani and
Paolo Pagano for useful discussions and suggestions.

This work is supported at INAF by funding from 
ASI on grant number I/R/039/04 and by COFIN MIUR
grant prot. number 2005025417,
at Penn State by NASA contract NASS5-00136 and 
at the University of Leicester by the Science and Technology Facilities Council. 
We gratefully acknowledge the contribution of dozens 
of members of the XRT team at OAB, PSU, UL, GSFC,
ASDC and our sub-contractors, who helped make 
this instrument possible.


\end{document}